\documentclass[12pt,preprint]{aastex}

\begin{document}

\title{The fractal dimension of star-forming regions\\
       at different spatial scales in M33}

\author{N\'estor S\'anchez\altaffilmark{1},
        Neyda A\~nez\altaffilmark{2},
        Emilio J. Alfaro\altaffilmark{1},
        Mary Crone Odekon\altaffilmark{3}}

\altaffiltext{1}{Instituto de Astrof\'{\i}sica de Andaluc\'{\i}a,
                 CSIC, Apdo. 3004, E-18080, Granada, Spain.}
\altaffiltext{2}{Departamento de F\'{\i}sica, Facultad de
                 Ciencias, Universidad del Zulia, Maracaibo, Venezuela.}
\altaffiltext{3}{Department of Physics, Skidmore College,
                 Saratoga Springs, NY 12866, USA.}

\email{nestor@iaa.es}

\slugcomment{\small\it The Astrophysical Journal: accepted}

\begin{abstract}
We study the distribution of stars, HII regions, molecular gas,
and individual giant molecular clouds in M33 over a wide range
of spatial scales. The clustering strength of these components
is systematically estimated through the fractal dimension.
We find scale-free behavior at small spatial scales and a
transition to a larger correlation dimension (consistent
with a nearly uniform distribution) at larger scales.
The transition region lies in the range $\sim 500-1000$ pc.
This transition defines a characteristic size that separates
the regime of small-scale turbulent motion from that of
large-scale galactic dynamics. At small spatial scales,
bright young stars and molecular gas are distributed with
nearly the same three-dimensional fractal dimension
($D_{f,{\rm 3D}} \lesssim 1.9$), whereas fainter stars
and HII regions exhibit higher values $D_{f,{\rm 3D}}
\simeq 2.2-2.5$. Our results indicate that the interstellar
medium in M33 is on average more fragmented and irregular
than in the Milky Way.
\end{abstract}


\section{Introduction}

In the Milky Way, gas and dust are organized in a hierarchical
and self-similar manner that it is supposed to be a consequence
of turbulent processes \citep{Elm04}. These fractal patterns are
observed over at least the range $0.1 \lesssim r \lesssim 100$ pc,
going from dense cores to giant molecular clouds
\citep[GMCs,][]{Ber07}. The formation of stars also shows fractal
features, usually observed in regions with a spatial hierarchy
ranging from a few pc up to about a kpc for so-called star
complexes \citep[e.g.][]{Efr95,Elm10}. Star clusters are in 
the lower levels of this hierarchy, although there is evidence
that young open clusters also exhibit smaller substructure
\citep{Sch08,San09}. 

This complexity can be characterized in many different ways. Some
common strategies, such as estimating mass or size distributions
of certain types of objects (clouds, cores, clusters), have to be
taken with extreme caution because they depend, among other things,
on the criteria adopted to define the objects \citep{Pin09,Cur10}.
Note that in a rigorously hierarchical scenario there is no
characteristic spatial scale that can be used to define any
particular structure. Moreover, it has been shown that projection
effects can significantly alter the estimated masses and sizes of
molecular clouds \citep{San06,She10}. Many other tools are now
widely used to describe the complexity of these structures
objectively and quantitatively \citep[see][for a comprehensive
review]{Elm04}. Fractal analysis is particularly appropriate for
dealing with hierarchical and self-similar systems. The fractal
dimension $D_f$, which quantifies the spatial heterogeneity,
can be calculated for both the distribution of gas and the
distribution of star-forming sites. This approach allows for
the comparison of similar objects in different regions and/or
under different physical conditions, and also the comparison
of different types of objects.

It is often accepted that the fractal dimension observed in the
interstellar medium (ISM) has a nearly universal value around
$D_f \simeq 2.3 \pm 0.3$ \citep[e.g.][]{Elm96}. This universality
would indicate either that interstellar turbulence is driven by
the same physical mechanisms everywhere, or that different physical
mechanisms can (and tend to) generate essentially the same type
of structures. However, the robustness of this conclusion is
questionable given the wide variety of results reported and
their associated uncertainties. If $D_f$ is inferred from
properties such as cloud masses or sizes, the resulting
uncertainties may be unacceptably large \citep{San06}.
A more appropriate strategy is to measure the fractal
dimension directly from an observed map. The boundaries
of the projected images of interstellar clouds in the Galaxy
have fractal dimension values spread over the range $1.2 \lesssim
D_{\rm per} \lesssim 1.5$ \citep{San05} but it is not clear whether
there are real variations from region to region or whether the
different values reflect different observational data and/or
analysis techniques. Moreover, even though $D_{\rm per} \simeq
1.3$ is a constant, the assumption that one can relate $D_{\rm
per}$ to the fractal dimension in three dimensions using
$D_f = D_{\rm per} + 1 \simeq 2.3$ has been shown not to
be valid \citep{San05}.

In a previous work we analyzed several emission maps of three
different molecular clouds (Ophiuchus, Perseus and Orion) and
obtained $D_f \simeq 2.6 \pm 0.1$ with no evidence of significant
variations \citep{San07b}. Similar fractal patterns should be
observable in the distribution of newborn stars if the distribution
of high-density cores follows the spatial structure of the parental
cloud. In fact, the young massive stars in the Gould Belt exhibit
a fractal pattern with the similar value $D_f = 2.68 \pm 0.04$
\citep{San07a}. However, a recent study by \citet{Sch10} reports
remarkable differences in the $\Delta$-variance spectra between
low-mass star forming clouds and massive GMCs. Star clusters also
show a wide variety of spatial patterns even for young, embedded
clusters \citep{Lad03,Car04,Sch08}, but it is not clear whether 
these variations are due to evolution or to differences in the
structure of the original clouds \citep{Goo04,Sch06,All09}.

Simulations of turbulent fluids produce very different structures
depending on which processes are considered in the system. For
example, \citet{Fed09} showed that simulations of supersonic 
isothermal turbulence in the extreme case of purely compressive
energy injection, produce a significantly smaller fractal dimension
for the density distribution ($D_f \sim 2.3$) than in the case of
purely solenoidal forcing ($D_f \sim 2.6$). Although it is widely
accepted that turbulence is the primary driver of the structure
and motion of the ISM, the main energy sources for this turbulence
are not yet well established. Part of the problem lies in the wide
variety of physical processes that can generate turbulent motions,
such as protostellar jets and outflows, (proto)stellar winds
and ionizing radiation, expanding HII regions, supernovae,
cloud collisions, galaxy interactions, and gravitational,
magnetorotational and other fluid instabilities \citep[recent reviews
on interstellar turbulence can be seen in][]{Elm04,Mac04,Bur06,Mck07}.
It is reasonable to expect that, depending on the dominant physical
mechanisms driving the turbulence, the resulting structure may differ
from region to region in the Galaxy.

A related issue is the spatial extent of this self-similar behavior,
since different physical mechanisms might dominate on different scales.
In the solar neighborhood, fractal behavior has been observed for the
distribution of young open cluster and young stars at spatial scales
of up to $\sim 1$ kpc \citep{Fue06,San07a,Fue09}. In external galaxies,
hierarchical structures extend up to $\gtrsim 1$ kpc scales for the gas
and for stars and star-forming sites \citep[some recent examples are
in][]{Bas07,Dut08,Gie08,Ode08,San08,Bas09,Dut09b,Sch09,Bon10}. Some
galaxies show a change in clustering strength with scale. This has 
been seen in both gas \citep{Elm01,Pad01,Kim07,Dut08,Dut09a} and
young stars \citep{Ode08}. There is also evidence for variations
among galaxies. Recent work indicates that the power spectrum of
the gas distribution is steeper for galaxies with a larger star
formation rate per unit area \citep{Wil05,Dut09b}. This trend
is consistent with the claim that bright galaxies tend to
distribute their star-forming sites in less clustered patterns
\citep{Par03,Ode06,San08}. In other words, a larger star formation
rate in a galaxy tends to be correlated with a larger fractal
dimension. Thus, at least on galactic spatial scales, a universal
picture for the fractal properties does not seem to hold.

A significant challenge in interpreting different measurements
is that authors present fractal dimensions for different ranges
of scales, identify their samples in different ways, and make
different assumptions regarding the comparison between
two-dimensional measures of clustering strength and the
implications for the actual three-dimensional structure.
Our approach here is to consider a case study in which we
systematically analyze the clustering of different components
of a single galaxy over a wide range of spatial scales. Because
of its proximity, large size, and low inclination, M33 is a
suitable object for this task. \citet{Bas07} showed that
the star formation in this galaxy appears to be inherently
hierarchical, with no characteristic size for star-forming
regions. However, \citet{Ode08} found a transition on a scale
of several hundred pc in the slope of the autocorrelation
function for young stars. In this work, we study the
clustering strength in the distribution of young stars,
HII regions, molecular gas, and individual giant molecular
clouds. Treating each of these components as consistently
as possible, we can directly compare the clustering strength
of each component as a function of scale. In addition, we
examine in detail the steepness of the transition in
clustering strength, evaluate the range of scales over
which it occurs, and analyze its meaning in the context
of the maximum spatial scale of coherent star formation.
We apply robust algorithms that have been previously
developed and tested on real data and simulated fractals.
Section~\ref{datos} describes the data and
Section~\ref{resultados} describes our calculation
of the fractal dimensions. An analysis and discussion
of the results are presented in Section~\ref{discusion}.
Finally, a summary and the main conclusions are given in
Section~\ref{conclusiones}.

\section{The data}
\label{datos}

The catalog of \citet{Mas06}, available via the
VizieR\footnote{http://vizier.u-strasbg.fr} database,
provides $UBVRI$ photometry of $146,622$ stars in M33
down to apparent magnitudes of 23 with photometric
errors less than 10\%. From this catalog we use the
same set of young stars (ages $\lesssim 30$ Myr)
that \citet{Ode08} used to calculate the angular
two-point autocorrelation function, that is, stars
fulfilling the criteria $-0.3 \leq V-I \leq 0.0$ and
$-6.0 \leq M_I \leq -4.0$. This set of stars was
divided into two subsets that we refer to simply
as ``bright" stars ($-0.3 \leq V-I \leq -0.1$ and
$-6.0 \leq M_I \leq -5.0$) and ``faint" stars
($-0.3 \leq V-I \leq 0.0$ and $-4.5 \leq M_I
\leq -4.0$). The total numbers of stars for
the bright and faint sets are $534$ and $1644$,
respectively. For the HII regions we use the catalog
of \citet{Hod99}, which gives the positions of $1272$
HII regions in M33. From this total data set, we remove
regions classified as unresolved, diffuse, linear and/or
any other factor that may raise doubts on the real nature
of such regions. We also remove the regions having null
integrated H$\alpha$ fluxes in the catalog. This means
that we only consider regions having an isophote level 
of at least $10^{-16}$ erg s$^{-1}$ cm$^{-2}$ arcsec$^{-2}$.
With these requirements the total number of ``bright" HII
regions to consider is $617$. The distribution of giant
molecular clouds in M33 is obtained from the catalog of
\citet{Ros07} (also available through VizieR) from which
we extract the positions of $149$ GMCs.

\begin{figure}[th]
\epsscale{1.0}
\plotone{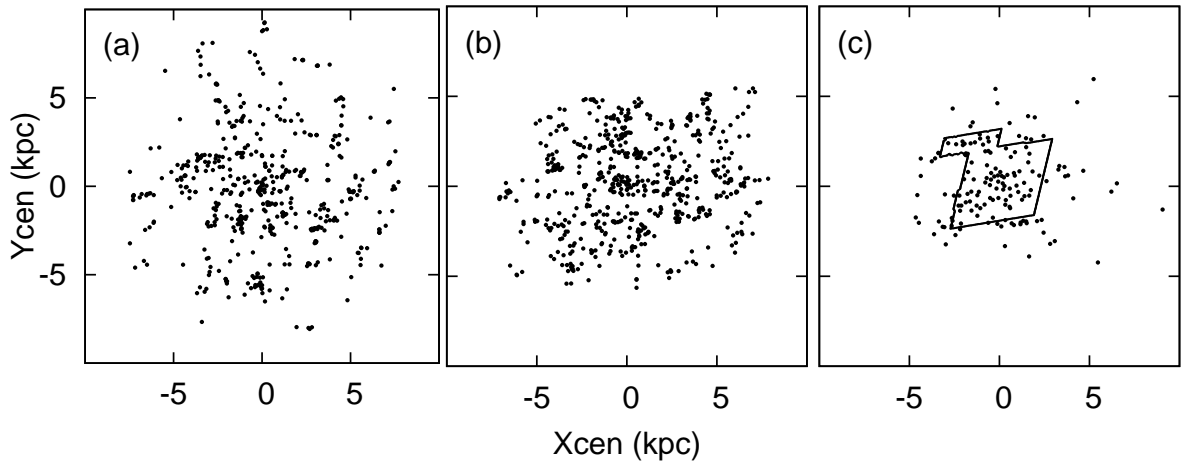}
\caption{\small Spatial distributions of (a) bright
stars, (b) bright HII regions and (c) GMCs in M33.
The axis coordinates are positions relative to the
galactic center in kpc. The inset in panel (c) shows
the area corresponding to the molecular gas map
(Fig.~\ref{mapaCO}).}
\label{distribuciones}
\end{figure}

We adopt a position angle of $23$ degrees and
an inclination of $55$ degrees (taken from
Hyperleda\footnote{http://leda.univ-lyon1.fr})
to deproject the positions of stars, HII regions
and GMCs. This is a first step necessary to avoid
the calculated fractal dimension becoming smaller
than its true value \citep{San08}. To convert the
angular sizes into linear sizes, we assume a distance
of $960$ kpc \citep{Bon06,U09}. The positions of bright
stars, HII regions and GMCs relative to the galactic center
$\alpha_{J2000.0} = 1^{\rm h} 33^{\rm m} 51^{\rm s}$,
$\delta_{J2000.0} = +30\arcdeg 39\arcmin 37\arcsec$)
are shown in Figure~\ref{distribuciones}.

\begin{figure}[th]
\epsscale{.5}
\plotone{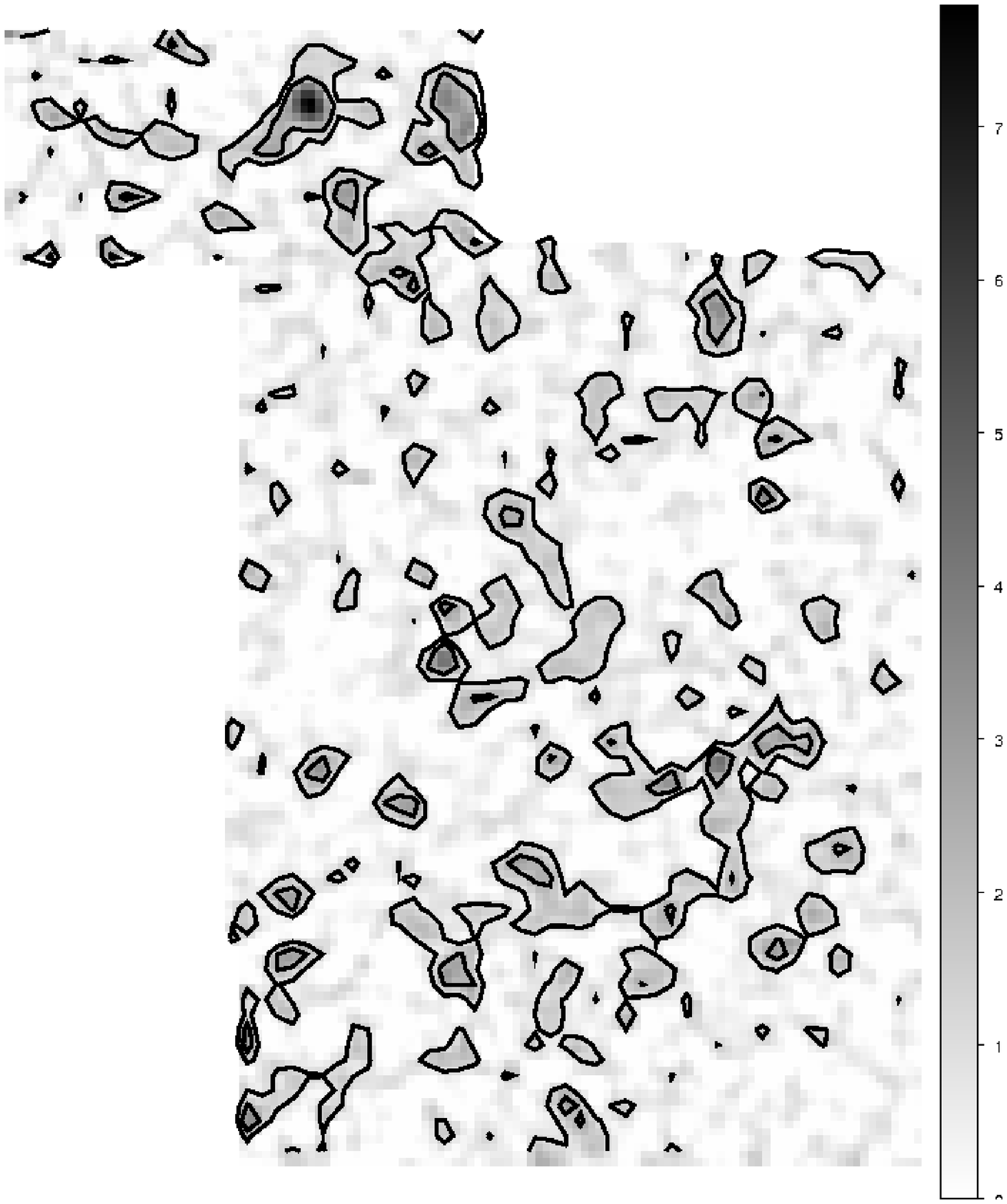}
\caption{\small Integrated intensity of CO emission in 
the central region of M33 (Fig.~\ref{distribuciones}).
The gray scale runs linearly from 0 (white) to the maximum
value (black, $\sim 7.8$ K km s$^{-1}$). Contour levels are
shown at $1$ and $2$ K Km s$^{-1}$.}
\label{mapaCO}
\end{figure}

As an additional way to study the the distribution of
molecular gas, we also directly use the high resolution
maps of CO ($J = 1 \rightarrow 0$) emission for the center
region of M33, kindly provided by Erik Rosolowsky \citep{Ros07}.
To avoid problems that might arise when estimating the fractal
properties in noisy maps \citep{San07b}, we use only the
combined NRO+BIMA+FCRAO data cube which has the highest
signal-to-noise ratio. The noise varies across the map but
the typical rms noise temperature is 60 mK \citep{Ros07}.
The final resolution is $20"$ ($93$ pc). We collapse the
data cube to produce the map of integrated intensity of CO
emission shown in Figure~\ref{mapaCO}.

\section{Estimation of the fractal dimension}
\label{resultados}

The degree of clustering of a point distribution can
be quantified by the correlation dimension $D_c$. For
scale-free sets, this can be calculated from $C(r)
\sim r^{D_c}$, where the correlation integral $C(r)$
is the average number of stars within a distance $r$
of each star. In practice, the distribution of stars
is not truly scale-free. For example, \citet{Ode08}
found a transition in the clustering strength of young
stars in M33 at a scale of about $300$ pc. Because the
correlation integral for any particular scale $r$ includes
information about the clustering on all scales less than
$r$, any deviation from a scale-free distribution will
affect, to some degree, the form of $C(r)$ on scales
larger than the deviation. For this reason, it is
sometimes preferable to use the differential form $c(r)$,
where $C(r) \sim \int_0^r c(r')dr'$, especially if the
goal is to determine the scale at which a change in
clustering strength takes place. Because of its
differential nature, however, $c(r)$ is too messy
for small data sets.  

Another reason the distribution of stars is not
actually scale-free is the finite size of a galaxy.
For an otherwise scale-free distribution with a
well-defined edge, one approach to dealing with
this limitation is to include, for each scale $r$,
only those stars at least $r$ from the edge.
\citet{San07a} used the minimum-area convex
polygon that contains all the points of finite 
two-dimensional fractal distributions, and showed
that it is an effective way of performing an edge
correction. We use this method to create an
edge-corrected version for $C(r)$. Finally,
on small scales the observed distribution of
stars is not scale-free because of resolution
limits and, ultimately, because of the size
of the stars themselves. \citet{San07a} found
that the calculation of $C(r)$ for two-dimensional
fractals is reliable on scales for which the standard
deviation of $C(r)$ is not greater than $C(r)$ itself.
We use this as our guide in determining the smallest
scales to include in our fits to the data.

\begin{figure}[th]
\epsscale{1.0}
\plotone{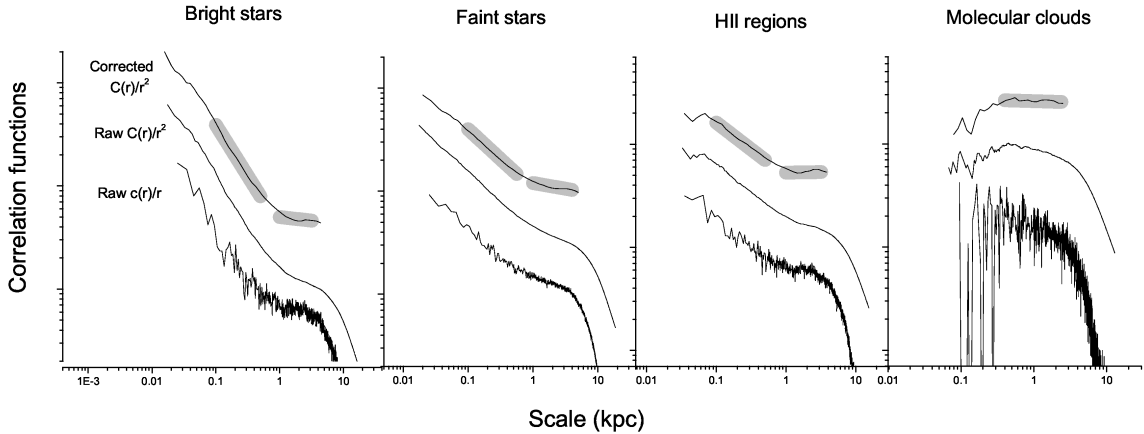}
\caption{\small From top to bottom in each panel, the
edge-corrected correlation integral $C(r)/r^2$, the raw
correlation integral $C(r)/r^2$, and the differential
correlation function $c(r)/r$ for the distributions of
pointlike objects. The data have been arbitrarily shifted
vertically for clarity. Shaded line segments show power-law
fits.}
\label{Threefunctions}
\end{figure}

Figure~\ref{Threefunctions} shows
our results for correlation integral $C(r)$ and the
differential form $c(r)$ for each type of point
distribution in our sample: bright stars, faint
stars, HII regions, and GMCs. Each panel includes
three functions; from top to bottom, they are the
edge-corrected correlation integral $C(r)/r^2$,
the uncorrected version of $C(r)/r^2$, and the
differential correlation function $c(r)/r$. The
functions are normalized so that a random
two-dimensional distribution (corresponding to
$D_c=2$) would produce a horizontal line. In
other words, the slope on the log-log plot is
$D_c-2$, so that a more strongly clustered
distribution decreases more quickly with scale.
A comparison of the bottom two functions in each
panel illustrates the fact that the differential
form $c(r)$ is more sensitive to changes with
scale, but messier, than the integral form
$C(r)$. A comparison of the top two function 
in each panel illustrates the effect of the
edge-correction algorithm, flattening the
slope on large scales.

Gray shaded line segments in Figure~\ref{Threefunctions}
show our power law fits to the edge-corrected $C(r)$.
Following \citet{San07a} and \citet{San08}, the range
in $r$ used for the fits is limited on small scales to
regions where the variation in $C(r)$ is not greater
than $C(r)$ itself;  these lower limits are $\sim 100$
pc for stars and HII regions and $\sim 450$ pc for GMCs.
In order to consider the possibility of a transition
in $D_c$, we performed separate fits at large and small
scales. The range of spatial scales for the transition was
varied and the result that minimizes the sum of the squared
residuals, is that shown in Figure~\ref{Threefunctions} and
listed in Table~\ref{tab01Dc}. We calculated uncertainties
in $D_c$, also listed in Table~\ref{tab01Dc}, using a
bootstrap algorithm.

All of the objects except the GMCs exhibit scale-free
clustering on small scales and a clear transition to a
flatter slope near $1$ kpc. The transition region is
found to lie in the range $500 \lesssim r \lesssim
1000$ pc for bright stars and $600 \lesssim r \lesssim
950$ for faint stars and HII regions.

\begin{figure}[th]
\epsscale{.7}
\plotone{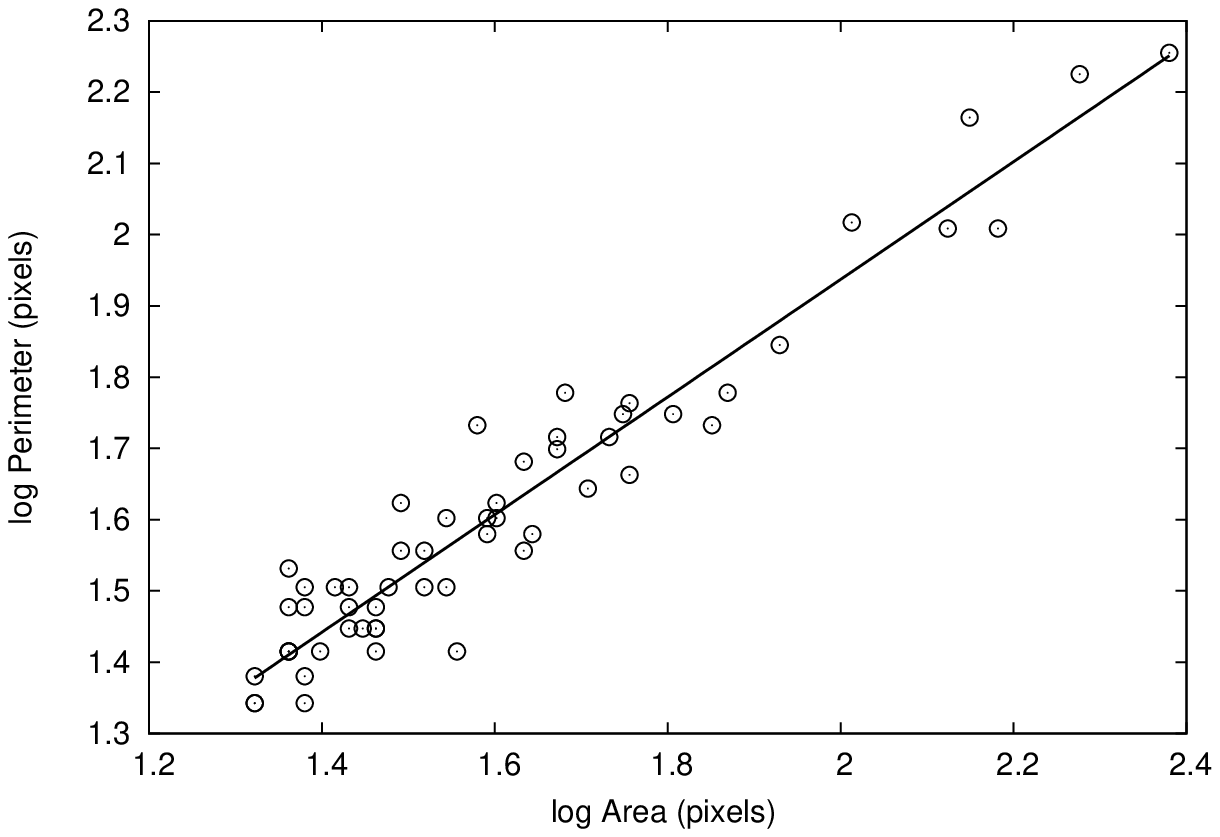}
\caption{\small Perimeter as a function of the area for
objects in the CO emission map (Fig.~\ref{mapaCO}) at
different intensity levels. The slope of the best linear
fit (solid line) is $0.825 \pm 0.032$.}
\label{figDper}
\end{figure}

We use a very different type of clustering measure on
the projected CO emission map (Fig.~\ref{mapaCO}).  
For these data, we calculated the perimeter-based
dimension $D_{\rm per}$ using the same algorithm
as in \citet{San05}. We chose brightness levels
from the lowest to the highest values in steps of
$0.5$ K Km s$^{-1}$. At each brightness level the
algorithm defines ``objects" in the image as sets 
of connected pixels whose brightness value is above
this level. Then, the perimeters $P$ and areas $A$
of each object are determined and the perimeter
dimension is obtained from the relation $P \sim
A^{D_{\rm per}/2}$. We did not consider objects 
containing less than $20$ pixels because most of
the structural details are lost in objects this
small \citep{San05}. For the same reason we also
excluded objects touching the edge of the map.
The perimeter-area log-log plot is shown in
Figure~\ref{figDper} and
the resulting perimeter dimension (twice the slope
of the best fit) is also listed in Table~\ref{tab01Dc}. 
The maximum spatial scale involved in this calculation
can be estimated as $\sqrt{A_{max}}$, being $A_{max}$
the area of the largest projected cloud in the map. We
obtained $A_{max}=240$ pixel$^2$ and the pixel size is
$6.87$ arcsec \citep{Ros07}, then the maximum size is
$\sim 500$ pc.


\begin{deluxetable}{lrcccccc}
\tablecolumns{8}
\tablewidth{0pt}
\tablecaption{Summary of Calculated Fractal Dimensions for M33
\label{tab01Dc}}
\tablehead{
\colhead{} &
\colhead{} &
\colhead{} &
\multicolumn{2}{c}{Small spatial scales\tablenotemark{a}} &
\colhead{} &
\multicolumn{2}{c}{Large spatial scales} \\
\cline{4-5} \cline{7-8} \\
\colhead{Sample} &
\colhead{$N_{dat}$} &
\colhead{} &
\colhead{$D_{f,{\rm 2D}}\tablenotemark{b}$} & \colhead{$D_{f,{\rm 3D}}$} & &
\colhead{$D_{f,{\rm 2D}}$} & \colhead{$D_{f,{\rm 3D}}$}
}
\startdata
Bright stars    & 534& & $1.01\pm0.05$ & 1.0-1.9 & & $1.93\pm0.03$ & 2.8-2.9 \\
Faint stars     &1644& & $1.42\pm0.04$ & 2.2-2.4 & & $1.89\pm0.02$ & 2.8-2.9 \\
HII regions     & 617& & $1.48\pm0.08$ & 2.3-2.5 & & $2.01\pm0.03$ & 2.9-3.0 \\
Molecular clouds& 149& & \nodata       & \nodata & & $1.98\pm0.04$ & 2.8-3.0 \\
CO emission map &\nodata & & $1.65\pm0.06$ & 1.6-1.8 & & \nodata & \nodata
\enddata
\tablenotetext{a}{For bright and faint stars and HII regions
small spatial scale means $\lesssim 500$ pc and large scale
means $\gtrsim 1$ kpc. For molecular gas large scale is
$\gtrsim 500$ pc (distribution of clouds) and small scale
is $\lesssim 500$ pc (CO map).}
\tablenotetext{b}{$D_{f,{\rm 2D}}$ refers either to the
two-dimensional correlation dimension $D_c$ (for the
distribution of stars, HII regions and GMCs) or to the
perimeter-area based dimension $D_{per}$ (for the CO map).
$D_{f,{\rm 3D}}$ is the corresponding three-dimensional
fractal dimension.}
\end{deluxetable}


Table~\ref{tab01Dc} summarizes the fractal dimensions calculated
for each type of data. The 2D fractal dimensions were converted
to 3D fractal dimensions using results from previous studies.
For the distribution of point objects (stars, HII regions and
GMCs) we used the results for projected disks derived in
\citet[Fig.~1]{San08} assuming a flatness (that is, a disk
thickness to diameter ratio) of $10^{-2}$ \citep{Ma98}. For
the CO map we used the results obtained in \citet{San05} with
a ``resolution" (i.e.  the maximum distance in the image in
pixel units) of $N_{pix} \sim 200$ pixels. In any case, the
corresponding 3D fractal dimension for the perimeter-area
dimension derived here is very weakly dependent on $N_{pix}$
\citep[see Fig.~8 in][]{San05}. The procedure of estimating
a 3D fractal dimension from its corresponding projected value
usually increases the associated uncertainties. We have included
in column labeled $D_{f,{\rm 3D}}$ in Table~\ref{tab01Dc} the
range of three-dimensional dimensions that are compatible with
the calculated values of $D_c$ or $D_{\rm per}$. It is important
to note, however, that these relatively large uncertainties do
not arise from the method used to calculate $D_f$, but from
possible random fluctuations in the projection of the original
fractals. That is, it is not possible in principle (at least
using only information from spatial distribution) to disentangle
which of the projected clumps are real and which are the result
of chance groupings during the projection.

\section{Discussion}
\label{discusion}

The first thing we note is a statistically significant
difference between the projected dimensions at relatively
small spatial scales and those at larger spatial scales. At
small scales the correlation dimension of the distribution
of stars and HII regions is $\lesssim 1.5$ whereas at large
scales it is $\gtrsim 1.9$, and the differences are always
larger than the associated uncertainties (Table~\ref{tab01Dc}).
The spatial scale where this transition takes place is roughly
the same for each component ($\sim 500-1000$ pc). The transition
from a smaller correlation dimension to a larger one for the young
stars in M33 was first reported by \citet{Ode08}. Here we provide
a more detailed quantification of this transition and observe
it for the first time for the distribution of HII regions.
The transition is not observed for the distribution of GMCs,
but given the limited number of data points for this sample
($N=149$) the lower limit of reliable values is higher than
for the other objects ($r \gtrsim 500$ pc).

What is the nature of this transition? It is not a simple edge
effect. Our algorithm corrects for edge effects and, as discussed
by \citet{Ode08}, we would expect an edge effect to cause a {\it
decrease} in the slope on larger scales. On the other hand, some
kind of change in the measured dimension is expected for projected
fractal disks. Given a monofractal of three-dimensional dimension
$D_{f,{\rm 3D}}$, the projected fractal dimension $D_{f,{\rm 2D}}$
of an extracted slice will depend on the slice thickness $H$
\citep[see][for detailed models and discussion]{San08}. For very
thin slices $D_{f,{\rm 2D}}=D_{f,{\rm 3D}}-1$ whereas for thick
slices $D_{f,{\rm 2D}} = \min\left\{ 2,D_{f,{\rm 3D}} \right\}$
\citep{Fal90}. Then, when calculating the correlation function we
should observe a change in the behavior around the scale $H$. For
$r \gg H$ the distribution can be seen as a thin slice whereas
$r \ll H$ corresponds to a very thick slice (this latter situation
is analogous to a simple, direct, projection). However, this is
expected to produce {\it smaller} values of $D_{f,{\rm 2D}}$ for
larger spatial scales \citep[see Fig.~1 in][]{San08}. This kind
of transition has been observed in the distribution of neutral
hydrogen in the LMC \citep{Elm01} and NGC~1058 \citep{Dut09a}. The
two-dimensional power spectra of these galaxies flattens at large
scales, indicating a smaller fractal dimension; this behavior can
be interpreted in terms of a change from 3D fluctuations on small
scales to 2D fluctuations on scales larger than the disk thickness.
In fact, power spectrum analysis has been suggested as a tool to
determine the thickness of nearly face-on galactic gas layer
\citep{Elm01,Pad01}. This effect contrasts with those presented
here for molecular gas and young stars, where the correlation
dimension is higher on larger scales. 

A different possibility was suggested by \citet{Pad01}. They
argued that there must be a {\it physical} transition in the
statistical properties of the flow close to the disk scale
height. In turbulent flow the energy is injected at certain
spatial scale and then it ``cascades" to smaller scales. But
there are many possible energy sources that may be relevant
at different levels---for instance, stellar outflows at
small scales and galactic shear at large scales. A possible
consequence may be different distribution patterns at
different size ranges. Even though the underlying turbulent
structure tends to be the same, non-turbulent motions acting
on galactic scales could modify the final structure at those
scales. In other words, the power law behavior at small spatial
scales would be a direct consequence of the self-similar turbulent
motions in the medium, but this turbulence is unlikely to extend to
very large scales, where two-dimensional flows should dominate the
dynamics. The transition from small-scale three-dimensional
turbulence to two-dimensional large-scale motions on the disk
should occur around the galactic scale height. Interestingly,
the behavior we observe is that all the fractal dimensions
at $r \gtrsim 1$ kpc are within a narrow range of values
($D_{f,{\rm 2D}} = 1.9-2.0$, or $D_{f,{\rm 3D}} = 2.8-3.0$)
that are consistent with essentially uniform (random)
distributions. Thus, we identify a characteristic spatial
scale (around $500-1000$ pc) that separates two different
physical regimes. This scale is of the order of the typical
size estimated by \citet{Efr98} for the largest star complexes
in spiral galaxies, which correlates with the galaxy isophotal
radius $R_{25}$ \citep{Elm83}. The transition zone separates
regions where coherent star formation is occurring in a
turbulent medium (stellar complexes) from larger regions
that are organized by large-scale galactic dynamics. The
analysis of the transition in the fractal dimension provides
an interesting method to determine the typical size of star
complexes in a given galaxy.

Despite the rather large uncertainties for $D_{f,{\rm 3D}}$,
it can be clearly seen that there are two separate range of
results for the three-dimensional fractal structures at small
spatial scales. On one hand, bright stars and molecular gas
are distributed with $D_{f,{\rm 3D}} \lesssim 1.9$. Young,
newborn stars should reflect the same conditions of the ISM
from which they were formed, which is roughly consistent with
this result. On the other hand, faint stars and HII regions
have $D_{f,{\rm 3D}} \simeq 2.2-2.5$. These higher fractal
dimension values can be interpreted in terms of evolutionary
effects. The clumpy distributions of young stars in LMC
\citep{Bas09} and SMC \citep{Gie08} evolve towards smoother
distributions as stellar ages increase. The same is true
for the stars clusters in these galaxies \citep{Bon10}.
It has been shown that the {\it brightest} HII regions in
spiral galaxies (which reflect, in a first approximation,
the initial distribution of star-forming sites) tend to be
distributed in more clumpy patterns than the low-brightness
regions \citep{San08}. The higher fractal dimensions for
faint stars and HII regions at small scales are probably
due to this kind of evolutionary effect.

We can compare the small-scale three-dimensional structure
of the interstellar medium in M33 with that of our own
galaxy. The spectral index $\gamma$ of the power spectrum
in the Milky Way lies in the range $2.8 \lesssim
\gamma \lesssim 3.2$ in 2D maps \citep{Elm04}. \citet{Bru02b}
systematically measured the energy spectrum in several
CO emission maps of the outer Galaxy and obtained $\beta =
2.17 \pm 0.31$, significantly higher than the Kolmogorov
spectrum $\beta = 5/3 \simeq 1.67$. The corresponding 
dimension of the projected contours is $D_{\rm per} \simeq
1.4$, very similar to the average value found from direct
measurements of $D_{\rm per}$ in the Galaxy \citep{Elm04}.
In a previous work \citep{San07b} we used various molecular
emission maps to study the fractal properties of Ophiuchus,
Perseus, and Orion, and found that the dimension is always
in the range $D_{\rm per} \simeq 1.30-1.35$. See Appendix~A
for a discussion of the definitions and relationships among
$\gamma$, $\beta$, and $D_{\rm per}$.

The 3D fractal dimension of the distribution of molecular gas
for M33 ($D_{f,{\rm 3D}} \simeq 1.6-1.8$, Table~\ref{tab01Dc}),
meanwhile, is much smaller than for the Milky Way
\citep[$D_{f,{\rm 3D}} \simeq 2.6-2.8$, see][]{San05,San07b}.
The CO emission map value of $D_{\rm per} = 1.65$ corresponds 
to $\beta = 1.7$, very close to the Kolmogorov value of
$5/3$\footnote{This is only a reference value. We are not
saying that turbulence in M33 actually behaves according
to Kolmogorov's description.}.
\citet{Elm03} found, from azimuthal scans of M33, that the power
spectrum of optical emission has an index even smaller than the
Kolmogorov value, although the uncertainties are relatively high.
That is, the boundaries of the projected perimeters in M33 are
on average more irregular (higher $D_{per}$ values) than those
of the Milky Way because molecular clouds in M33 exhibit a more
fragmented structure (smaller $D_{f,{\rm 3D}}$ values). The fact
that $D_{f,{\rm 3D}} \lesssim 1.9$ for the three-dimensional
distribution of very young stars is consistent with the small
fractal dimension of the CO gas. In addition, \citet{Bas07},
using the minimum spanning tree technique, found that the
number of objects increases with the radius as $N \sim
R^{\alpha}$ with $\alpha=1.63-1.96$. Random sampling in
perfect hierarchical fractals yields $N \sim R^{D_f}$ so
that, in general, their results are consistent with the
relatively small fractal dimensions we are reporting here.
Whether there are (or not) other differences
with GMCs in our Galaxy remain an open problem.
\citet{Eng03} found that the mass spectrum of GMCs in M33
follows a power-law distribution (according to a hierarchical
picture) but with a slope considerably steeper than that found
in the Milky Way. However, \citet{She08} concluded that if the
same cloud identification algorithm and analysis technique are
applied, and if resolution-dependent effects are taken into
account, then there seem to be no differences in the GMCs
properties between these two galaxies. \citet{Bol08}
analyzed in a consistent manner GMC properties (sizes,
velocity dispersions, and luminosities) in several galaxies
(including M33) subject to a wide variety of physical
conditions and found only small differences with the
properties of GMCs in the Milky Way.
However, it is important to keep in mind the difficulty in
comparing observations of an external galaxy with observations
of the Milky Way. Molecular clouds from large-scale surveys in
the Milky Way can be blended in space and velocity, making it
troublesome to extract information on their internal structures.
Blending effects are reduced in external galaxies such as M33
that are seen nearly face-on. Moreover, the internal structure
of GMCs in the Milky Way is measured for clouds in the solar
neighborhood, at an average Galactocentric distance of
about $8$ kpc. Instead, the average internal structure
of clouds in M33 was calculated in this work for the
molecular map of the central region, at galactocentric
distance less than $\sim 2-3$ kpc.
Obviously, more detailed studies are needed to clarify
this point but, in this work, we have found a significant
and remarkable difference in the internal structure of GMCs
between M33 and the Milky Way.

\section{Conclusions}
\label{conclusiones}

In this work we measure the clustering in the distribution
of stars, HII regions, molecular gas, and individual giant
molecular clouds in M33 as a function of spatial scale. We
identify a transition in the clustering strength from a
very clumpy (fractal) structure at relatively small scales
to almost uniform patterns at larger scales. The spatial
scale for this transition is in the range $\sim 500-1000$
pc, independently from the type of object considered. We
argue that this characteristic scale separates two physical
regimes, one where small-scale turbulent motions generate
self-similar structures and another dominated by large-scale
galactic dynamics.

The existence of this transition implies that care must taken
in calculating a single value for the fractal dimension over
a large range of scales. One must take into account not only 
edge effects on large scales and sampling/resolution effects
on small scales, but also the possibility of physical changes
on intermediate scales. Ideally, this could be done using the
differential form of the correlation function combined with
an edge-correction algorithm; this way, changes with scale
are not carried into the calculation of the dimension for
larger scales. However, for small data sets (even the larger
data sets presented here, with on the order of $1000$ objects),
the differential form is rather noisy. 

At small spatial scales, bright stars and molecular gas in
M33 are distributed with nearly the same three-dimensional
fractal dimension. This result is consistent with the idea
that newborn stars follow the same patterns of the ISM from
which they were formed. Faint stars and HII regions exhibit
higher fractal dimensions possibly as a consequence of
evolutionary relaxation.

Interestingly, the three-dimensional fractal dimension of
molecular gas in M33 ($D_{f,{\rm 3D}} \lesssim 1.9$) is
significantly smaller than in the Milky Way ($D_{f,{\rm 3D}}
\simeq 2.7 \pm 0.1$). This means that the interstellar
medium in M33 is on average more fragmented and irregular
than in our own galaxy. This result may have important
implications for the study of the physical processes
that determine the ISM structure.

\acknowledgments
We acknowledge Erik Rosolowsky for providing the
CO emission maps of the central region of M33. We
want to thank Bruce Elmegreen for helpful comments
and discussion. We also thank an anonymous referee
for his/her comments.
This research has made use of NASA's Astrophysics
Data System and of HyperLeda and VizieR databases.
We acknowledge financial support from MICINN of Spain
through grant AYA2007-64052 and from Consejer\'{\i}a
de Educaci\'on y Ciencia (Junta de Andaluc\'{\i}a)
through TIC-101 and TIC-4075. N.S. is supported by
a post-doctoral JAE-Doc (CSIC) contract.
E.J.A. acknowledges financial support from the
Spanish MICINN under the Consolider-Ingenio 2010
Program grant CSD2006-00070: ``First Science with
the GTC".

\appendix

\section{Appendix}
\label{apendice}

Interstellar turbulence can be characterized in several
different ways. Two commonly used measures are the energy
spectrum $E(k) \sim |k|^{-\beta}$ and the power spectrum
$P(k) \sim |k|^{-\gamma}$ for wavenumber $k$. The energy
spectrum is the integral over all directions of the power
spectrum, $E(k)dk = \int P(k) d\Omega$, so that with this
nomenclature the scaling exponents $\beta$ and $\gamma$
are related through \citep{Bru02a}
\[
\gamma = \beta + E -1 \ \ ,
\]
where $E$ is the euclidean dimension of the image. For
example, for a dissipationless cascade of energy through
an incompressible fluid the Kolmogorov energy spectrum is
$\beta=5/3$ and the power spectrum index is $\gamma = 11/3$,
$8/3$, or $5/3$ in 3D, 2D or 1D, respectively \citep{Elm04}.

The structure of molecular clouds can also be
described as a fractional Brownian motion (fBm)
structure \citep{Stu98,Miv03}. The properties of
fBm clouds depend on a single parameter, the so-called
Hurst exponent $H$. This parameter characterizes the
self-similar structure and it is related to the corresponding
power spectrum through \citep{Miv03}
\[
\gamma = 2H+E \ \ .
\]
The iso-intensity contours of the $E$-dimensional
image of a fBm-cloud is given by $D_{\rm iso}=E-H$ \citep{Stu98},
which corresponds to the perimeter-area based dimension $D_{\rm per}$
for two-dimensional maps. In this last case ($E=2$) we can write the
equations relating these commonly used parameters as
\[
D_{\rm per} = 2-H \ \ ,
\]
\[
D_{\rm per} = (5-\beta)/2 \ \ ,
\]
and
\[
D_{\rm per} = (6-\gamma)/2 \ \ .
\]
For very rough structures ($H=0$), we have $\beta=1$,
$\gamma=2$ and $D_{\rm per}=2$, while for very smooth
structures ($H=1$) we get $\beta=3$, $\gamma=4$ and
$D_{\rm per}=1$.

\end{document}